\documentclass[12pt]{aipproc}

\title{Matrix Distributed Processing and FermiQCD}

\author{Massimo Di Pierro}

\affiliation{Fermilab, Batavia, IL 60510, USA}

\date{31st October 2000}

\begin{document}

\maketitle

\begin{abstract}
Matrix Distributed Processing is a collection of classes and functions
written in C++ for fast development of efficient parallel algorithms for
the most general lattice/grid application. FermiQCD is an Object Oriented Lattice QCD
application of MDP, under development at Fermilab.

\end{abstract}

\section{Introduction}

It is believed that, down to the smallest observed length scale, fundamental interactions 
in nature are local. This means that the equations one writes to describe the 
physical world are, in the majority of cases, local differential equations or systems of local 
differential equations. They can be non-linear, strongly coupled and stochastic but, if they 
describe a fundamental interaction, they are also local.

With very few exceptions, these equations do not have an exact analytical solution, therefore they must be solved numerically. This is done by discretizing the space on which the equations are defined and applying iteratively the appropriate algorithm.

The most general local differential equation contains derivatives which, 
after discretization, becomes quasi-local terms. For example
\begin{equation}
\frac{\textrm{d}^n}{\textrm{d}x} \phi (x) \rightarrow \sum_{k=0}^{n} \frac{(-1)^k}{(2a)^n} \left(\!\! \begin{tabular}{c}
k \\ n 
\end{tabular} \!\!
\right) \phi(x+(n-2k)a)
\label{discretize}
\end{equation}
where $a$ is the lattice spacing introduced in the discretization process.
``Quasi-local'' here means that a local term (the $n$-th derivative in $x$) becomes 
a linear combination of non-local terms localized within a radius $n a$ form $x$.

The typical iterative algorithm that solves a local differential equation has the 
form 
\begin{equation}
\textrm{ITERATE}\hskip 1cm \forall x: \ \ \ \phi(x) = H(x,\phi(y))
\label{iter}
\end{equation}
where $H$ is some function of the position, $x$, and of the value of the field, $\phi(y)$, 
in some neighborhood of $x$ within $|x-y| \le n a$.

The exact form of $H$ is not completely determined by initial differential equation, since there 
are different inequivalent ways to discretize it. A difference in the discretization procedure 
means a difference in the convergence speed and a difference in the discretization errors 
(that vanish with $a \rightarrow 0$).

Finding the numerical solution can be very costly but these algorithms 
can be very efficiently parallelized 
(using a supercomputer and/or a cluster of workstations).
This is because one can partition the space $x$ on which the field $\phi(x)$ is defined over 
different CPUs. Each CPU applies the algorithm, eq.~\ref{iter}, to the local sites and 
this can be done in parallel. Because of the quasi-locality of the function $H$ 
it is necessary that each process maintains an updated copy of the field 
variables $\phi(y)$ for each $y$ in the neighborhood of the local sites $x$.
Each CPU will distinguish between local sites $\{x\}$ (the sites stored by the CPU), boundary sites $\{y\}$ (sites that are not local but a local copy exists because they must be accessed) and hidden sites (sites that do not affect the computation performed by that particular CPU).

For every parallel algorithm to work it is necessary to keep the boundary sites updated, 
i.e. if a field variable at a particular site is modified by one of the CPU, 
its copies (maintained by different CPUs) have to be modified accordingly.
This requires communication among the different CPUs.

Matrix Distributed Processing (MDP)~\cite{me1} provides the tools to implement this kind of 
algorithms on a computer in an easy and object oriented way. It also provides 
some basic classes for matrix manipulations, statistical analysis and a random number generator.

Communications in MDP are based on Message Passing Interface which is {\it de facto} a standard 
for parallel applications. MPI calls are hidden inside the basic classes that constitute 
MDP and are invisible to the user.

\section{Example}

As a first example of an application, let us consider here the 
following problem: 

{\bf Problem:}\ {\it Solve numerically, in $U$, the following equation 
\begin{equation}
\nabla ^ 2 U = \cos(U+V)
\label{eq1}
\end{equation}
where $U(x)$ and $V(x)$ are fields of $3 \times 3$ matrices defined on a four 
dimensional space $x$ with the topology of a torus $T^4$. $V(x)$ is initialized 
with random $SU(3)$ matrices.} (In this example $U(x)$ plays the role of the field 
$\phi(x)$ of the last section.)

{\bf Solution:} The first step is to discretize the space on which the 
fields are defined by approximating it with a $N^4$ 
lattice (with $N=8$). The second step consists in writing 
down a discretized form of eq.~\ref{eq1}, using eq.~\ref{discretize}.
 In adimensional units (defined by 
imposing $a=1$) one obtains
\begin{eqnarray}
U(x) &=& H(x,U) \nonumber \\ 
 &\equiv& \frac18 \big[ 
\cos(U(x)+V(x))+\nonumber \\
&& \hskip 7mm U(x+\hat 0)+U(x-\hat 0)+ \nonumber \\
&& \hskip 7mm U(x+\hat 1)+U(x-\hat 1)+\nonumber  \\
&& \hskip 7mm U(x+\hat 2)+U(x-\hat 2)+\nonumber \\
&& \hskip 7mm U(x+\hat 3)+U(x-\hat 3)
\big] + {\mathcal O}(a)
\label{eq2}
\end{eqnarray}
where $x\pm\hat n$ is $y=(x_0\pm\delta_{0,n},x_1\pm\delta_{1,n},
x_2\pm\delta_{2,n},x_3\pm\delta_{3,n})$.

The third and usually non-trivial step is writing a computer program 
that implements, in a parallel way, the recursive relation of eq.~\ref{eq2}.

Here is how this can be implemented using MDP:
\begin{verbatim}
01: #include "MDP_Lib2.h"
02: #include "MDP_MPI.h"
03: int main(int argc, char **argv) {
04:   mpi.open_wormholes(argc, argv);
05:   int box[4]={8,8,8,8};
06:   generic_lattice space(4,box);
07:   Matrix_field U(space,3,3);
08:   Matrix_field V(space,3,3);
09:   site x(space);
10:   forallsites(x) {
11:     U(x)=0;
12:     V(x)=space.random(x).SU(3);
13:   };
14:   U.update();
15:   V.update();
16:   for(int i=0; i<100; i++) {
17:     forallsites(x)
18:       U(x)=0.125*(cos(U(x)+V(x))+
19:                   U(x+0)+U(x-0)+
20:                   U(x+1)+U(x-1)+
21:                   U(x+2)+U(x-2)+
22:                   U(x+3)+U(x-3));
23:     U.update();
24:   };
25:   V.save("V_field.dat");
26:   U.save("U_field.dat");
27:   mpi.close_wormholes();
28:   return 0;
29: };
\end{verbatim}
\begin{itemize}
\item lines 1,2 read the MDP libraries;
\item lines 4 and 27 open and close the communication channels among the parallel processes;
\item line 6 defines the object {\tt space} belonging to the class 
{\tt generic\_lattice} with size specified by the {\tt box}; (by default a generic lattice has the topology of a torus but the user can specify a different topology. The user can also specify on which processor each lattice site is stored. MDP optimizes the communications accordingly)
\item lines 7,8 define the two fields of matrices {\tt U(x)} and {\tt V(x)};
\item line 9 defines a variable {\tt x} of class {\tt site} defined on the {\tt space};
\item lines 10-13 initialize the fields in parallel;
\item lines 14,15 take care of the communication to update the copies of the boundary sites;
\item lines 16-24 perform 100 iterations of the algorithm, eq.~\ref{eq2};
each iteration is automatically parallelized over the available CPUs;
\item lines 25,26 save the input and output fields.
\end{itemize}

Many lattice/grid problems can be solved in a similar way. MDP provides some of built-in 
field classes and the user can easily define its own field class which inherit the standardized 
{\tt update}, {\tt load} and {\tt save} member functions.
The standard {\tt load/save} functions guarantee the portability of data different platforms (both parallel and non-parallel).

MDP also features a parallel random number generator, i.e. one random generator 
for each lattice site, that insures reproducibility of computations
independently on the way the lattice is partitioned.

\section{FermiQCD @ Fermilab}

Fermilab is using MDP to develop a general purpose Object 
Oriented Lattice QCD application~\cite{me2}, 
called FermiQCD\footnote{%
FermiQCD can be downloaded from: \\
{\tt http://thpc16.fnal.gov/fermiqcd.html}
}.
The typical problem in QCD (Quantum Chromo Dynamics) is that of determining the correlation 
functions of the theory as function of the parameters. From the knowledge of these correlation
functions one can extract hadron masses and matrix elements and compare them with experimental 
results. This provides both a useful check of the theory (QCD in particular) and also a unique 
way to extract some of the fundamental parameters of the Standard Model 
(for example the CKM matrix elements). 

On the lattice, each correlation function is computed numerically as the average
of the corresponding operator applied to elements of a Markov chain 
of gauge field configurations.
Both the processes of building the Markov chain and of measuring 
operators involve quasi-local algorithms.

Some of the main features of FermiQCD are the following:

\begin{itemize}
\item it supports an arbitrary number of lattices in each parallel program and an arbitrary number of fields defined on each lattice;
\item each lattice can have an arbitrary dimension, arbitrary topology and arbitrary partitioning;
\item some of the basic built-in fields are: 

{\tt gauge\_field}, 

{\tt fermi\_field}, 

{\tt staggered\_field},

{\tt scalar\_field};

\item {\tt gauge\_field}s are in the adjoint representation of $SU(N_c)$ for an arbitrary $N_c$.
\end{itemize}

The basic parallel algorithms implemented in FermiQCD are~\cite{rothe}:
\begin{itemize}
\item heathbath Monte Carlo to create the Markov chain of gauge field configurations;
\item $O(a^2)$ improved heathbath Monte Carlo;
\item minimum residue inversion and stabilized biconjugate gradient
inversion for the fermionic matrix;
\item ordinary and stochastic fermionic propagators;
\item ordinary fermionic actions: Wilson, Clover ($O(a)$ improved) and D234 ($O(a^2)$ improved);
\item staggered fermionic actions: Kogut-Susskind, Lepage ($O(a^2)$ improved).
\end{itemize}

Moreover FermiQCD is able to read existing Lattice QCD data 
in the CANOPY/ACPMAPS format, in the UKQCD format and 
in the MILC format.

Here are few examples of FermiQCD Object Oriented capabilities 
(compared with examples in the standard textbook notation for Lattice QCD)

\noindent 1) {\bf QCD:} (algebra of Euclidean gamma matrices)
\begin{equation}
A=\gamma^{\,\mu} \gamma^{\,5} e^{3 i \gamma^{\,2}}
\end{equation}

\noindent {\bf FermiQCD:} 
\begin{verbatim}
Matrix A;
A=Gamma[mu]*Gamma5*exp(3*I*Gamma[2]);
\end{verbatim}

\noindent 2) {\bf QCD:} (multiplication of a fermionic field for a spin structure)
\begin{equation}
\forall x: \ \ \ \chi(x)=(\gamma^{\,3}+m) \psi(x + \hat \mu)
\end{equation}

\noindent {\bf FermiQCD:} 
\begin{verbatim}
/* assuming the following definitions
generic_lattice space_time(...);
fermi_field chi(space_time,Nc);
fermi_field psi(space_time,Nc);
site x(space_time);
*/
forallsites(x)
   chi(x)=(Gamma[3]+m)*psi(x+mu);
\end{verbatim}

\noindent 3) {\bf QCD:} (translation of a fermionic field)
\begin{equation}
\forall x,a: \ \ \ \chi_a(x)=U(x,\mu)\psi_a(x+\hat \mu)
\end{equation}

\noindent {\bf FermiQCD:} 
\begin{verbatim}
forallsites(x)
  for(a=0; a<psi.Nspin; a++)
    chi(x,a)=U(x,mu)*psi(x+mu,a);
\end{verbatim}

\section*{acknowledgments}

I wish to acknowledge the University of Southampton (UK) where the MDP project started 
and to thank the following members of the Fermilab Theory Group: E.~Eichten, J.~Juge, 
A.~Kronfeld, P.~MacKenzie and J.~Simone, for many suggestions and comments about FermiQCD.
I also acknowledge I borrowed many Lattice QCD algorithms from existing 
CANOPY, MILC and UKQCD programs; I thank here the authors for letting 
me study their codes.
\vskip 1mm
The project FermiQCD is supported by the U.S. Department of Energy under 
contract No. DE-AC02-76CH3000.

\end{document}